\documentclass[10pt]{article}
\usepackage{multicol}
\usepackage{graphicx}
\usepackage{amsmath}
\usepackage[a4paper]{geometry}
\usepackage{rotating}
\usepackage{verbatim}
\usepackage{hyperref}

\begin{document}

\begin{center}
{\Large\bf Statistical Behavior of Lepton Pair Spectrum in
Drell-Yan Process and Signal from Quark-Gluon Plasma in High
Energy Collisions}

\vskip.75cm

Xu-Hong Zhang{\footnote{E-mail: xhzhang618@163.com;
zhang-xuhong@qq.com}}, Fu-Hu Liu{\footnote{Correspondence E-mail:
fuhuliu@163.com; fuhuliu@sxu.edu.cn}}

\vskip.25cm

{\small\it Institute of Theoretical Physics \& Collaborative
Innovation Center of Extreme Optics \& State Key Laboratory of\\
Quantum Optics and Quantum Optics Devices, Shanxi University,
Taiyuan 030006, China}
\end{center}

\vskip.5cm

{\bf Abstract:} We analyze the transverse momentum ($p_{T}$)
spectra of lepton pairs ($\ell\bar \ell$) generated in the
Drell-Yan process, as detected in proton-nucleus (pion-nucleus)
and proton-(anti)proton collisions by ten collaborations over a
center-of-mass energy ($\sqrt{s_{NN}}$ or $\sqrt{s}$ if in a
simplified form) range from $\sim20$ GeV to above 10 TeV. Three
types of probability density functions (the convolution of two
L\'{e}vy-Tsallis functions, the two-component Erlang distribution,
and the convolution of two Hagedorn functions) are utilized to fit
and analyze the $p_{T}$ spectra. The fit results are approximately
in agreement with the collected experimental data. Consecutively,
we obtained the variation law of related parameters as a function
of $\sqrt{s}$ and invariant mass ($Q$). In the fit procedure, a
given L\'{e}vy-Tsallis (or Hagedorn) function can be regarded as
the probability density function of transverse momenta contributed
by a single quark ($q$) or anti-quark ($\bar q$). The Drell-Yan
process is then described by the statistical method.
\\

\\


\vskip1.0cm

\begin{multicols}{2}

{\section{Introduction}}

There are more than one processes that can generate a pair of
charged leptons ($\ell\bar \ell$) in experiments of high energy
collisions. In 1970, Sidney Drell and Tung-Mow Yan firstly
proposed $\ell\bar \ell$ production in a high energy hadron
scattering in a process we now call ``Drell-Yan"
process~\cite{Drell-Yan}. In this process, a quark ($q$) in one
hadron and an anti-quark ($\bar q$) in another hadron are
annihilated and a virtual photon ($\gamma^*$) or $Z$ boson is
generated, which then decays into $\ell\bar \ell$. This process is
expressed as $A+B\longrightarrow\gamma^{*}/Z+X\longrightarrow
\ell+\bar\ell+X$, where $A$ and $B$ are collision hadrons and $X$
denotes other particles produced in the collisions. The Drell-Yan
process has been extensively studied experimentally,
theoretically, and phenomenologically.

The literature about the theoretical description of the Drell-Yan
process within Quantum Chromodynamics (QCD) is well known and
settled~\cite{1a,1b,1c,1d,1e,1f,1g,1h}. The framework for the
description of the transverse momentum dynamics [sometimes,
indicated as Collins-Soper-Sterman (CSS) formalism] is summarized
in the well known book by John Collins~\cite{1a}. Some recent
reviews on the subject of transverse momentum distributions in
Drell-Yan process can be found in refs.~\cite{1c,1d,1e} in which
many works were cited. More theoretical works at both small and
large $p_T$ are available in literature~\cite{1f,1g,1h}. At the
same time, lots of phenomenological works were published in the
past~\cite{7,8,9,10,11,12}, recent~\cite{13,14,15,16,17,18}, and
very recent years~\cite{19,20,21,22}.

Several phenomenological interpretations of experimental Drell-Yan
data collected in the previous many years have been released by
various groups, particularly in recent years where first
extractions of quark transverse momentum distributions are
becoming available from highly accurate theoretical descriptions
of QCD perturbative ingredients. The factorization theorem for the
Drell-Yan process allows to write the
transverse-momentum-differential cross section as a convolution of
two transverse-momentum-dependent (TMD) parton distribution
functions (PDFs), which are, under certain conditions, very
complicated. This complicated factorization involves soft factors
that resum soft gluon radiation regularizing a certain class of
divergences that arise in the theoretical formulae. The soft gluon
resummation is especially important in the description of the
quark-gluon plasma (QGP), where $\ell\bar \ell$ can be produced in
a process similar to that of Drell-Yan but with different origin
of quarks.

QGP is a new form of matter which is created in the central region
of high energy nucleus-nucleus collisions, where extreme density
and high temperature environment is developed. It has become one
of the important areas of research in the field of nuclear and
particle physics. The gradual maturity of QCD and gauge field
theory provide a powerful explanation for this novel matter and
phenomenon. In fact, QGP is particularly short-lived. In QGP, a
quark $q$ and anti-quark $\bar q$ can soon be annihilated into a
virtual photon $\gamma^*$ or $Z$ boson, which then decays to a
pair of leptons $\ell\bar \ell$. This happens in the QGP
degeneration process in which most particles are produced. The
yield, invariant mass, rapidity ($y$), and transverse momentum
($p_T$) distribution of $\ell\bar \ell$ depend on the momentum
distribution of $q\bar q$ and gluons in QGP in the collision
region. Therefore, the information of $\ell\bar \ell$ can be used
to judge whether QGP is generated and further to study its
thermodynamic status making the $\ell\bar \ell$ production one of
the most important signals generated by QGP. Consequently, the
study on $\ell\bar \ell$ becomes particularly critical.

From the above, it is clear that $\ell\bar \ell$ can be produced
in high-energy hadronic/nucleonic collisions two main ways: the
Drell-Yan process and QGP degeneration process. To study the
properties of QGP, we should remove the influence of Drell-Yan
process and vice-versa. Generally, we may use the same methodology
to describe the two processes. At present, one mainly uses the
statistical method to study the properties of QGP.
Correspondingly, we may also use the statistical method to study
the production of $\ell\bar \ell$ in Drell-Yan process, especially
because the factorization theorem is very hard to model. In short,
the statistical description for Drell-Yan process is necessary to
better understand the properties of QGP.

The measurement of lepton-pair physical quantities (including
energy, $p_T$, $y$, etc.) in experiments studying the Drell-Yan
process provide lots of valuable information about the dynamic
properties and evolution process of the produced particles. In
particular, $p_T$ is Lorentz invariant in the beam direction and
can be used to describe the particles' motion and system's
evolution. There are different functions that can be used to
describe the $\ell\bar \ell$ $p_T$ spectra in statistics. For
example, we can use the L\'{e}vy-Tsallis
function~\cite{Levy-Tsallis1, Levy-Tsallis2, Levy-Tsallis,
Tsallis, Levy Tsallis}, the (two-component) Erlang
distribution~\cite{Erlang distribution1, Erlang distribution2,
Erlang distribution3}, and the Hagedorn function~\cite{Hagedorn1,
Hagedorn2} to fit the experimental data to obtain the analytical
parameters of the $p_T$ spectrum. Since the Drell-Yan process is
the result of the interactions of $q$ and $\bar q$, we can use the
convolution of two functions to describe the $p_T$ spectra. The
idea of convolution is concordant to the factorization theorem for
Drell-Yan process.

In this paper, we use three functions to fit and analyze the
Drell-Yan $\ell\bar \ell$ $p_T$ spectra obtained by ten
collaborations from the experiments of high energy proton-nucleus
(pion-nucleus) and proton-(anti)proton collisions. These
experimental studies provide a great resource for us to better
understand the collision mechanism and dynamic characteristics of
the mentioned process.
\\

{\section{Formalism and method}}

Naturally, the $p_T$ spectra of Drell-Yan $\ell\bar \ell$ depend
on collision energy. For that reason we should use different
probability functions to study these spectra at different
energies. Here we briefly describe the three functions which will
be used in this study. In the following, $p_{t1}$ and $p_{t2}$ are
the transverse momenta of the two quarks, and $p_T$ is the
transverse momentum of the two quark system, which equals the
transverse momentum of the dilepton system at leading order.
\\

{\subsection {The L\'{e}vy-Tsallis function}}

The Boltzmann distribution is the most important probability
density function in thermodynamic and statistical physics. We
present the probability density function of $p_T$ as a simple
Boltzmann distribution~\cite{Boltzmann-Gibbs1, Boltzmann-Gibbs2,
Boltzmann-Gibbs3}:
\begin{align}
f_{p_{T}}(p_{T})=\frac{1}{N}\frac{dN}{dp_{T}}=
C_{B}p_{T}\mathrm{exp}\left(-\frac{\sqrt{p_{T}^{2}+m_{0}^{2}}}
{T_{B}} \right),
\end{align}
where $N$ is the number of identical particles of mass $m_0$
produced in the collisions, $C_{B}$ is the normalization constant,
and $T_{B}$ is the effective temperature of the collision system.

The Boltzmann distribution is a special form of the Tsallis
distribution, and the latter has a few alternative
forms~\cite{Levy-Tsallis1,Levy-Tsallis2,Levy-Tsallis,Tsallis,Levy
Tsallis}. As one of the Tsallis distribution and its alternative
forms, the L\'{e}vy-Tsallis function of the $p_{T}$ spectrum of
hadrons~\cite{Levy-Tsallis1,Levy-Tsallis2,Levy-Tsallis,Tsallis,Levy
Tsallis} is used in this work. We have the following form to
describe the transverse momentum ($p_t$) distribution of
(anti-)quark:
\begin{align}
f_{1}(p_{t})=N_{q}\sqrt{p_{t}}\left[1+\frac{1}{nT}\left
(\sqrt{p_{t}^{2}+m_{q}^{2}}-m_{q}\right)\right]^{-n},
\end{align}
where $N_{q}$ is the normalization constant, $T$ and $n$ are the
fitted parameters, and $m_{q}$ is the mass of $q$ or $\bar q$
taking part in the reaction. In general, we use
$m_u=m_d=0.3~\mathrm{GeV}/c^{2}$ in Drell-Yan process because $q$
or $\bar q$ is from the participant hadrons. The same $m_u$ or
$m_d$ is for sea quarks and those in baryons, where the sea quarks
of higher mass are not considered in this work. In QGP,
$m_u=0.003~\mathrm{GeV}/c^{2}$ and $m_d=0.007~\mathrm{GeV}/c^{2}$
because the quarks are approximately bare~\cite{39}. It has been
verified that the Tsallis distribution is just a special case of
the L\'{e}vy distribution, but not the opposite~\cite{Levy
Tsallis}.
\\

{\subsection{The (two-component) Erlang distribution}}

The Erlang distribution~\cite{Erlang distribution1,Erlang
distribution2,Erlang distribution3} is proposed to fit the $p_T$
spectra in the multi-source thermal model~\cite{multisource
thermal model}. Generally, a two-component Erlang
distribution~\cite{Erlang distribution1,Erlang
distribution2,Erlang distribution3} is used to describe both the
soft and hard processes. The contribution fractions of the two
components are determined by fitting the experimental data. The
numbers of parton sources participating in the soft and hard
processes are represented by $n_{S}\geq2$ and $n_{H}=2$
respectively. The contribution ($p_t$) of each parton source to
$p_T$ of final-state particle is assumed to obey an exponential
function:
\begin{align}
f_{i}(p_{t})=\frac{1}{\langle p_{t}
\rangle}\mathrm{exp}\left(-\frac{p_{t}}{\langle
p_{t}\rangle}\right),
\end{align}
where $\langle p_{t}\rangle$ represents the average $p_t$
contributed by the $i$-th source. Because $\langle p_{t}\rangle$
is the same for different sources, the index $i$ in $\langle
p_{t}\rangle_i$ is omitted.

The $p_T$ distribution contributed by $n_S$ ($n_H$) sources is the
convolution of $n_S$ ($n_H$) exponential functions, which gives
the Erlang distribution. Let $k$ denote the contribution fraction
of the first component (soft process). The two-component Erlang
distribution is:
\begin{align}
f(p_{T})=&\frac{kp_{T}^{n_{S}-1}}{(n_{S}-1)!\langle
p_{t}\rangle_{S}^{n_{S}}}\mathrm{exp}\left( -\frac{p_{T}}{\langle
p_{t}\rangle}_{S}\right) \nonumber\\
&+\frac{(1-k)p_{T}}{\langle
p_{t}\rangle_{H}^{2}}\mathrm{exp}\left( -\frac{p_{T}}{\langle
p_{t}\rangle}_{H}\right).
\end{align}
Fitting the data with the two-component Erlang distribution, we
can get the changes of parameters $\langle p_{t} \rangle _{S}$,
$\langle p_{t}\rangle_{H}$, and $k$.

We should discuss the values of $n_S$ and $n_H$ further. If
$n_S>2$, the participant partons are expected to be $q\bar q$ and
$n_S-2$ gluons in the soft or non-violent annihilation process.
Considering that the probability of multi-parton participating
together in the process is low, we have usually $n_S=2$ or 3 in
this work. Generally, the larger the $n_S$, the sharper the
distribution peak. In many cases, $n_S=3$, means that $q\bar q$
and a gluon participate in the soft process. For all cases,
$n_H=2$ (always true in this work) means that only $q\bar q$
participate in the violent annihilation in the hard process.
\\

{\subsection {The Hagedorn function}}

The Hagedorn function is an inverse power
law~\cite{Hagedorn1,Hagedorn2} which is an empirical formula
derived from perturbative QCD. Generally, this function can only
describe the spectra at large $p_T$, but not the entire $p_T$
interval. In the case of using the Hagedorn function in a wide
range of $p_T$, the probability density function of $p_{T}$ can be
expressed as:
\begin{align}
f_1(p_T)=Ap_{T}\left( 1+\frac{p_{T}}{p_{1}} \right)^{-n_1},
\end{align}
where $A$ is the normalization constant and $p_{1}$ and $n_1$ are
the fitted parameters. The final-state particles with high momenta
are mainly produced by the hard scattering process during the
collisions. However, both the soft and hard processes contribute
to the $p_T$ spectra. In some cases, the soft excitation process
in the low $p_{T}$ range can also be described by the Hagedorn
function. We try to use the Hagedorn function to describe the
transverse momentum distribution of (anti-)quarks. That is, we may
use $p_t$ instead of $p_T$ in Eq. (5) to obtain the transverse
momentum distribution of (anti-)quarks, which is a new form of Eq.
(5) and will be used in the following section.

We have tested the Hagedorn function with different revisions in
which $p_T/p_1$ in Eq. (5) is replaced by $p_T^2/p_1^2$, $Ap_T$ is
replaced by $Ap_T^2/m_T$, or $p_T/p_1$ is replaced by
$p_T^2/p_1^2$ and $Ap_T$ is replaced by $A$, where $p_1$ and $A$
vary in different revisions. The uses of the revised Hagedorn
functions result in some over-estimations in low (or high) $p_T$
region comparing to the Hagedorn function. Contrarily, these
revisions result in some under-estimations in high (or low) $p_T$
region due to the normalization. The revisions of the Hagedorn
function are beyond the focus of the present work, we shall not
discuss them further.
\\

{\subsection {The convolution of functions}}

The convolution of functions is an important operation process in
functional analysis that can be used to describe the weighted
superposition of input and system response (that is, two
sub-functions). The Drell-Yan process is the result of the
interactions of $q$ and $\bar q$ in high energy collisions, which
means that we need the convolution of two functions to describe
this process. Indeed, the above Eq. (2) or (5) can be used to
describe the transverse momentum distribution, $f_1(p_{t1})$, of a
single (anti-)quark's contribution. The second (anti-)quark's
contribution is $f_2(p_{t2})=f_1(p_T-p_{t1})$, where $p_T$ is
still the transverse momentum of $\ell\bar \ell$ system. So the
convolution of two probability density functions should be used to
describe the $p_T$ spectrum of $\ell\bar \ell$ in the Drell-Yan
process. We have the convolution of two Eq. (2) or (5) to be
expressed as:
\begin{equation}
f(p_{T})=\int^{p_{T}}_0f_1(p_{t1})f_1(p_{T}-p_{t1})dp_{t1},
\end{equation}
where $f_1(p_{t1})$ [$f_1(p_{T}-p_{t1})$] is shown as Eq. (2) if
we use the L\'{e}vy-Tsallis function or Eq. (5) if we use the
Hagedorn function.

It should be noted that the total transverse momentum before (of
the two quarks system) and after (of the two leptons system) are
equal. The assumption that the total transverse momentum is equal
to the sum of the scalar transverse momenta of the two partons is
for the particular case in which the vectors
$\overrightarrow{p}_{t1}$ and $\overrightarrow{p}_{t2}$ are
parallel. Our recent works~\cite{41,42} show that this assumption
is in agreement with many data. Naturally, we do not rule out
other assumptions such as the particular case in which
$\overrightarrow{p}_{t1}$ and $\overrightarrow{p}_{t2}$ are
perpendicular and the general case which shows any azimuth for
$\overrightarrow{p}_{t1}$ and $\overrightarrow{p}_{t2}$. The
particular case used in this work is more easier than other
particular or general case in the fit to data. We are inclined to
use the parallel case.

As a legitimate treatment, the convolution formula Eq. (6) can be
used to fit the $p_T$ spectrum of $\ell\bar \ell$ in Drell-Yan
process, where $f_1(p_{t1})$ and $f_1(p_{T}-p_{t1})$ are from
empirical guess which is simpler than the factorization theorem
based on perturbative QCD. On one hand, Eq. (6) can reflect the
weighted contribution of the transverse momentum of each
(anti-)quark to the $p_T$ spectrum in the process. On the other
hand, Eq. (6) can also reflect the system in which two main
participants take part in the interactions. Using the convolution
to fit the data is a good choice for us, which allows us to more
accurately understand the interaction process and mechanism
between interacting partons, more completely describe the energy
dependence and interdependence of the function parameters, and
further better analyze the $p_T$ spectrum.
\\

\begin{figure*}[htbp]
\begin{center}
\includegraphics[width=12.0cm]{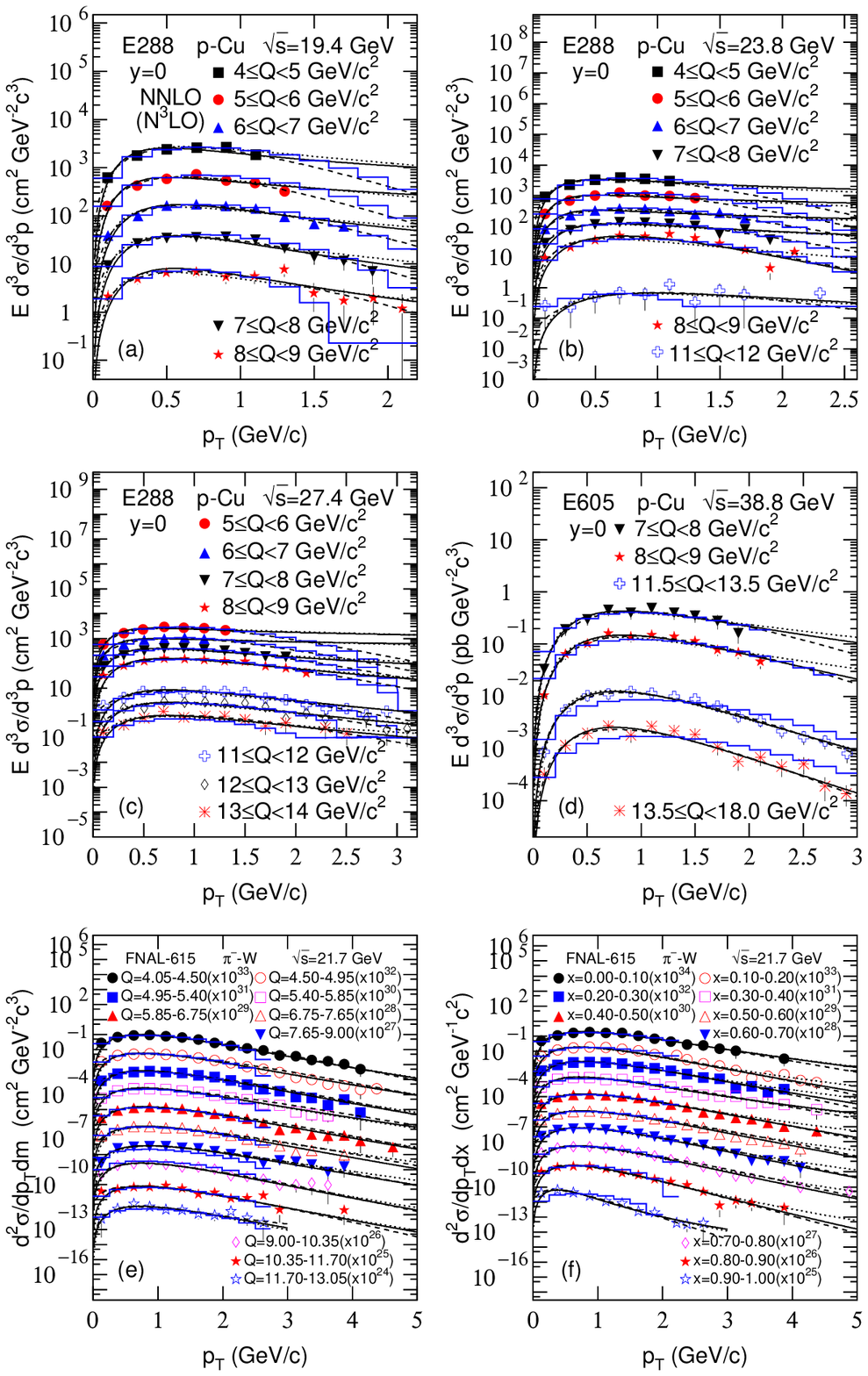}
\end{center} \vskip-2.5mm
{\small Figure 1: Transverse momentum spectra of $\ell\bar \ell$
(with different invariant masses $Q$ or Feynman variables $x_F$)
produced by the Drell-Yan process in different collisions at
different energies. The data points in Figures 1(a)--1(c) are
quoted from the E288 Collaboration~\cite{E288} and obtained in
$p$-Cu collisions at $\sqrt{s}=19.4$~GeV ($4\leq
Q<9~\mathrm{GeV}/c^{2}$), 23.8~GeV ($4\leq
Q<13~\mathrm{GeV}/c^{2}$), and 27.4~GeV ($5\leq
Q<13~\mathrm{GeV}/c^{2}$), respectively. The data points in Figure
1(d) come from $p$-Cu collisions at $\sqrt{s}=38.8$ GeV performed
by the E605 Collaboration~\cite{38.8}. The data points in Figures
1(e) ($4.05\leq Q<13.05~\mathrm{GeV}/c^{2}$) and 1(f) ($0.0\leq
x_F <1.0$) come from $\pi^-$-W collisions at $\sqrt{s}=21.7$~GeV
measured by the FNAL-615 Collaboration~\cite{21.7}, where the
units GeV/$c^2$ are not shown in Figure 1(e) due to crowding
space. In panel (d), $\ell\bar \ell$ are $e^+e^-$; and in other
panels, $\ell\bar \ell$ are $\mu^+\mu^-$. The solid, dashed, and
dotted curves are our results of fitting the data points with the
convolution of two L\'{e}vy-Tsallis functions [Eqs. (2) and (6)],
the two-component Erlang distribution [Eq. (4)], and the
convolution of two Hagedorn functions [Eqs. (5) and (6)],
respectively. See text in subsection 3.3 for the meanings of
histograms which are quoted from QCD calculations.}
\end{figure*}

\begin{figure*}[!htb]
\begin{center}
\includegraphics[width=12.0cm]{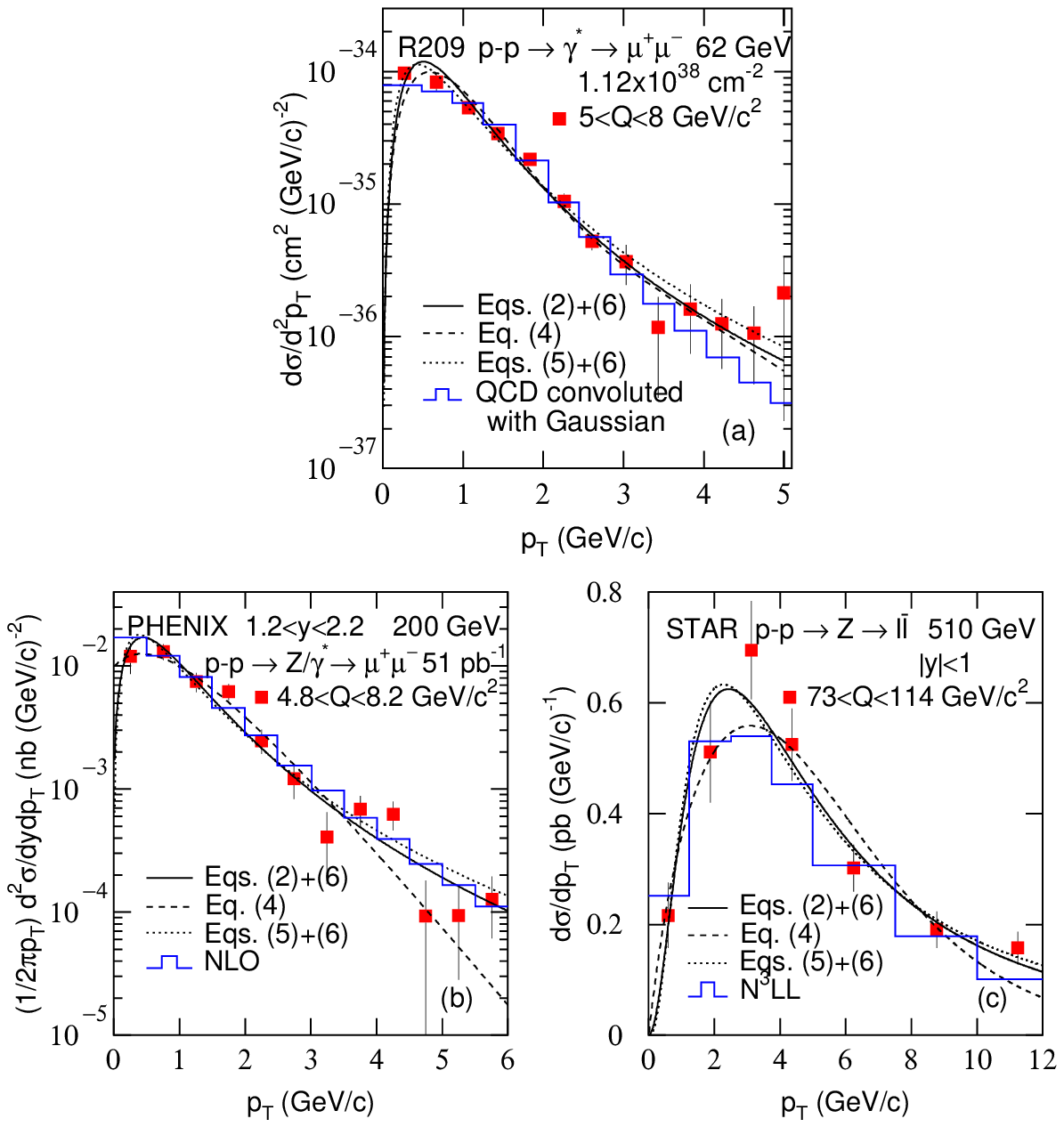}
\end{center}
{\small Figure 2: Transverse momentum spectra of $\ell\bar \ell$
(with different invariant masses $Q$) produced by the Drell-Yan
process in $pp$ collisions at $\sqrt{s}=62$ (a), 200 (b), and 510
GeV (c). The data points are quoted from the R209 (a)~\cite{62a},
PHENIX (b)~\cite{200}, and STAR Collaborations (c)~\cite{510}. In
panels (a) and (b), $\ell\bar \ell$ are $\mu^+\mu^-$. The solid,
dashed, and dotted curves are our results of fitting the data
points with the convolution of two L\'{e}vy-Tsallis functions
[Eqs. (2) and (6)], the two-component Erlang distribution [Eq.
(4)], and the convolution of two Hagedorn functions [Eqs. (5) and
(6)], respectively. See text in subsection 3.3 for the meanings of
histograms which are quoted from QCD calculations.}
\end{figure*}

\begin{figure*}[!htb]
\begin{center}
\includegraphics[width=12.0cm]{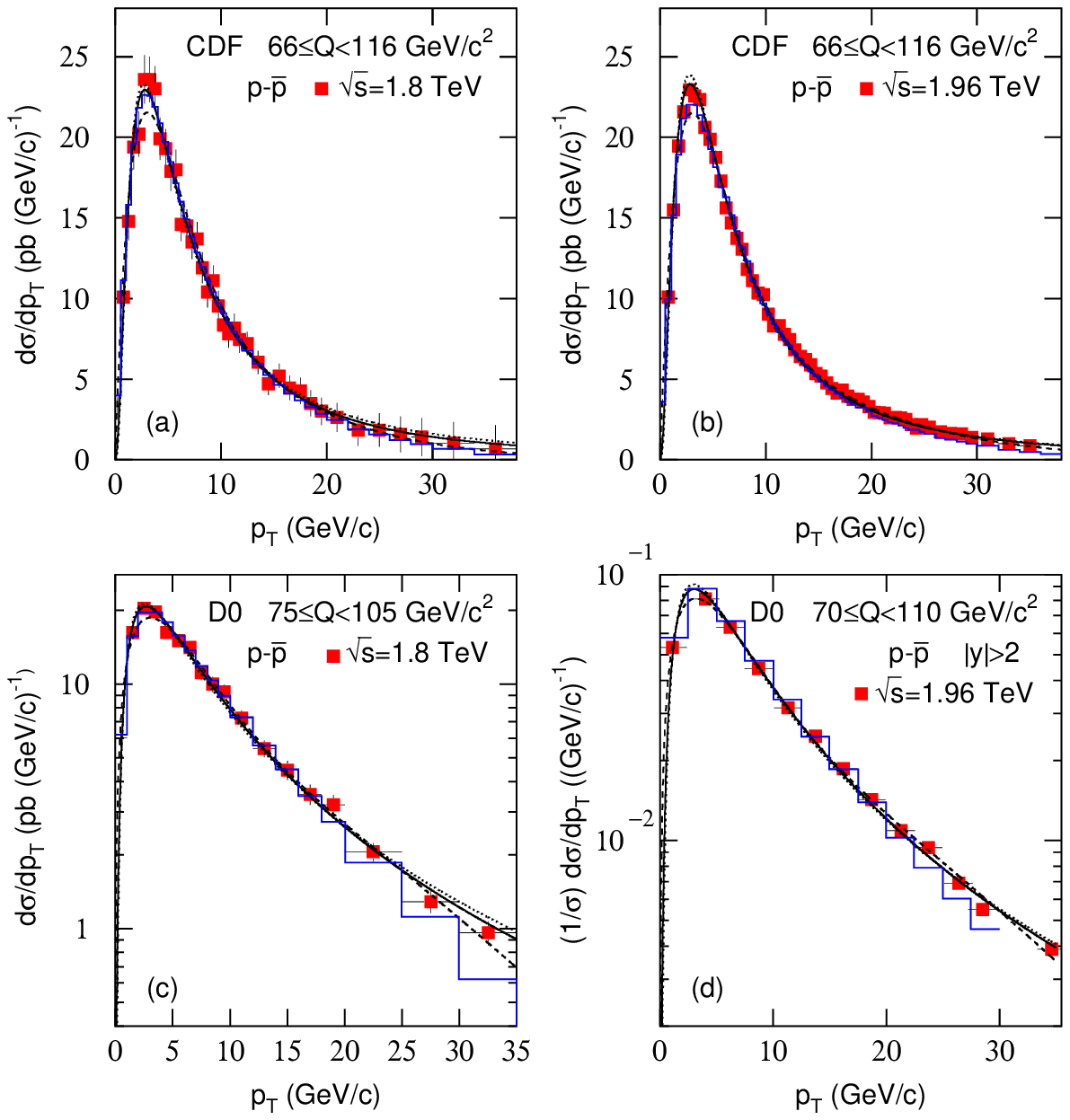}
\end{center}
{\small Figure 3: Same as Figure 2, but showing the $p_T$ spectra
of $\ell\bar \ell$ (with different invariant masses $Q$) produced
by the Drell-Yan process in $p\bar p$ collisions at $\sqrt{s}=1.8$
(a, c) and 1.96 TeV (b, d). The data points are quoted from the
CDF (a, b)~\cite{CDF1.8,CDF1.96} and D0 Collaborations (c,
d)~\cite{D01.8,D01.96-1,D01.96-2}. In all panels, $\ell\bar \ell$
are $e^+e^-$. See text in subsection 3.3 for the meanings of
histograms which are quoted from QCD calculations.}
\end{figure*}

\begin{figure*}[htbp]
\begin{center}
\includegraphics[width=12.0cm]{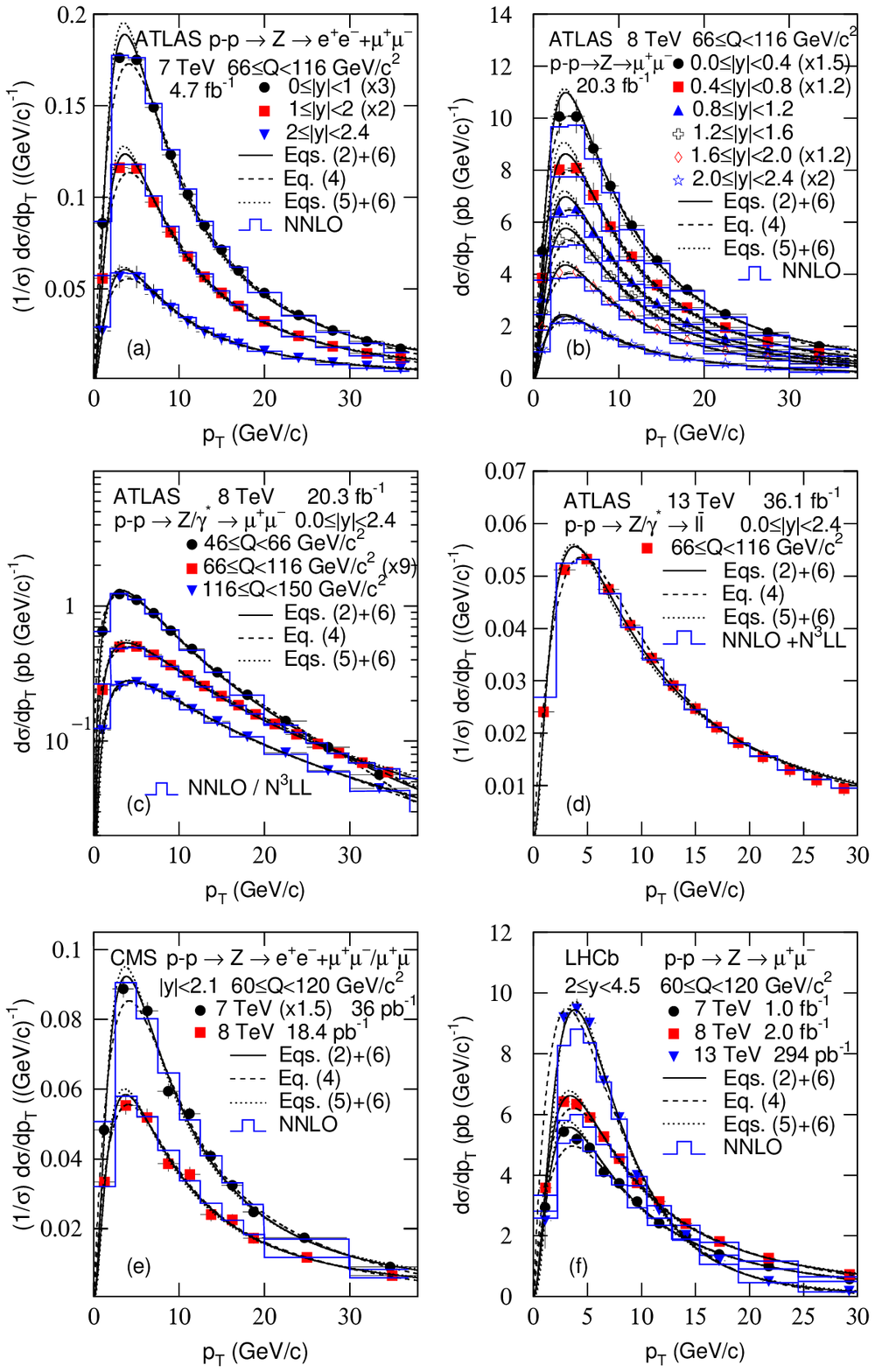}
\end{center} \vskip-2.5mm
{\small Figure 4: Same as Figure 2, but showing the $p_T$ spectra
of $\ell\bar \ell$ [with different conditions ($\sqrt{s}$, $Q$,
and $y$)] produced by the Drell-Yan process in $pp$ collisions at
the LHC energies. The data points are quoted from the experiments
performed by the ATLAS (a--d)~\cite{ATLAS7,ATLAS8,ATLAS13}, CMS
(e)~\cite{CMS7,CMS8}, and LHCb Collaborations
(f)~\cite{LHCb7,LHCb8,LHCb13}. The flavor of $\ell\bar \ell$ is
shown in the panels. See text in subsection 3.3 for the meanings
of histograms which are quoted from QCD calculations.}
\end{figure*}

\begin{figure*}[!htb]
\begin{center}
\includegraphics[width=12.0cm]{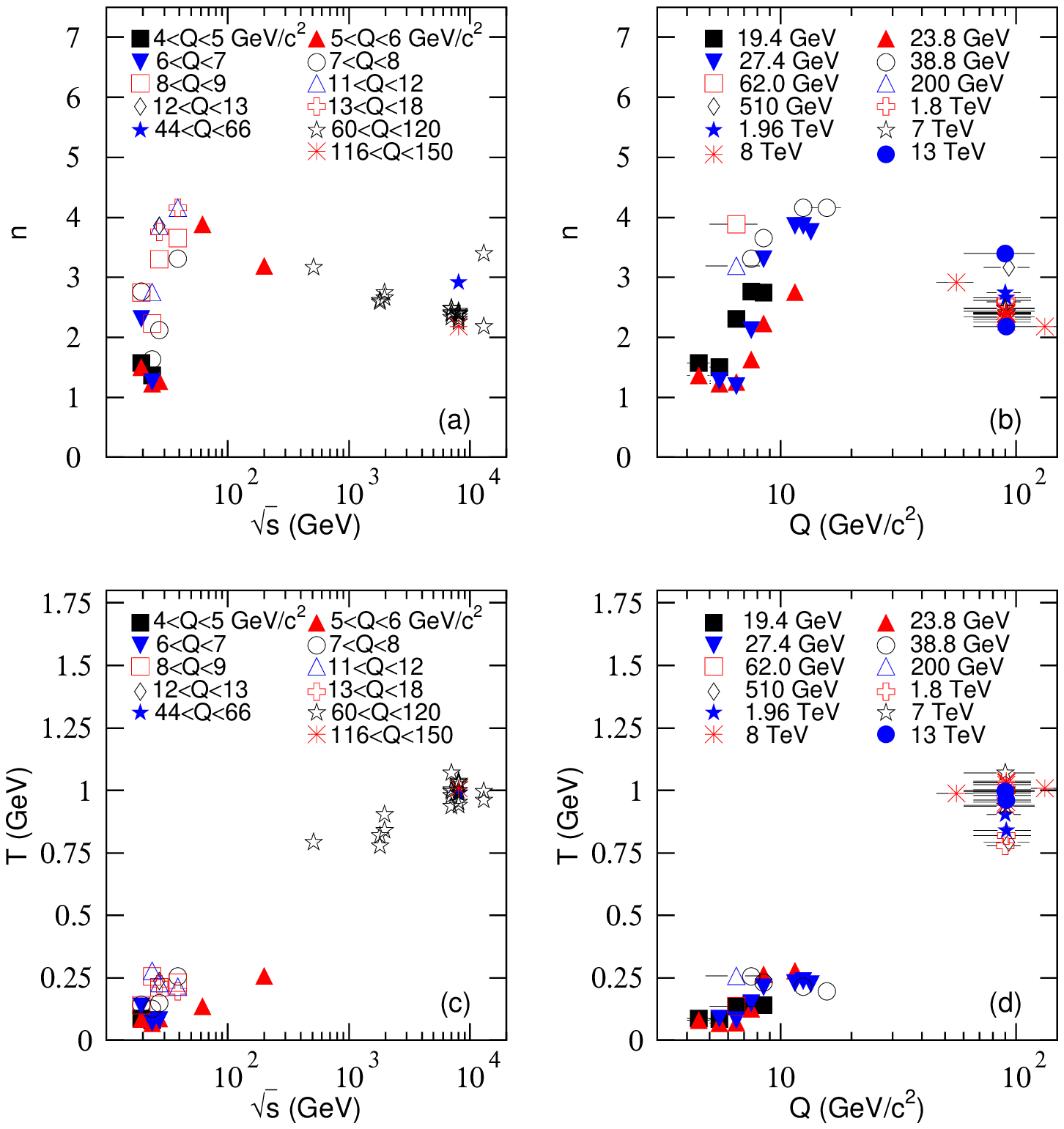}
\end{center}
{\small Figure 5: The variation of parameters $n$ (a, b) and $T$
(c, d) in the L\'{e}vy-Tsallis function with energy $\sqrt{s}$ (a,
c) and invariant mass $Q$ (b, d). The parameter values are taken
from Figures 1--4 and recorded in Tables 1 and 2. Since the
invariant mass grouping in Figure 1(e) is different from others,
and the varying quantity in Figure 1(f) is the Feynman variable,
Figure 5 does not include the parameters from Figures 1(e) and
1(f) to avoid trivialness.}
\end{figure*}

\begin{figure*}[htbp]
\begin{center}
\includegraphics[width=12.0cm]{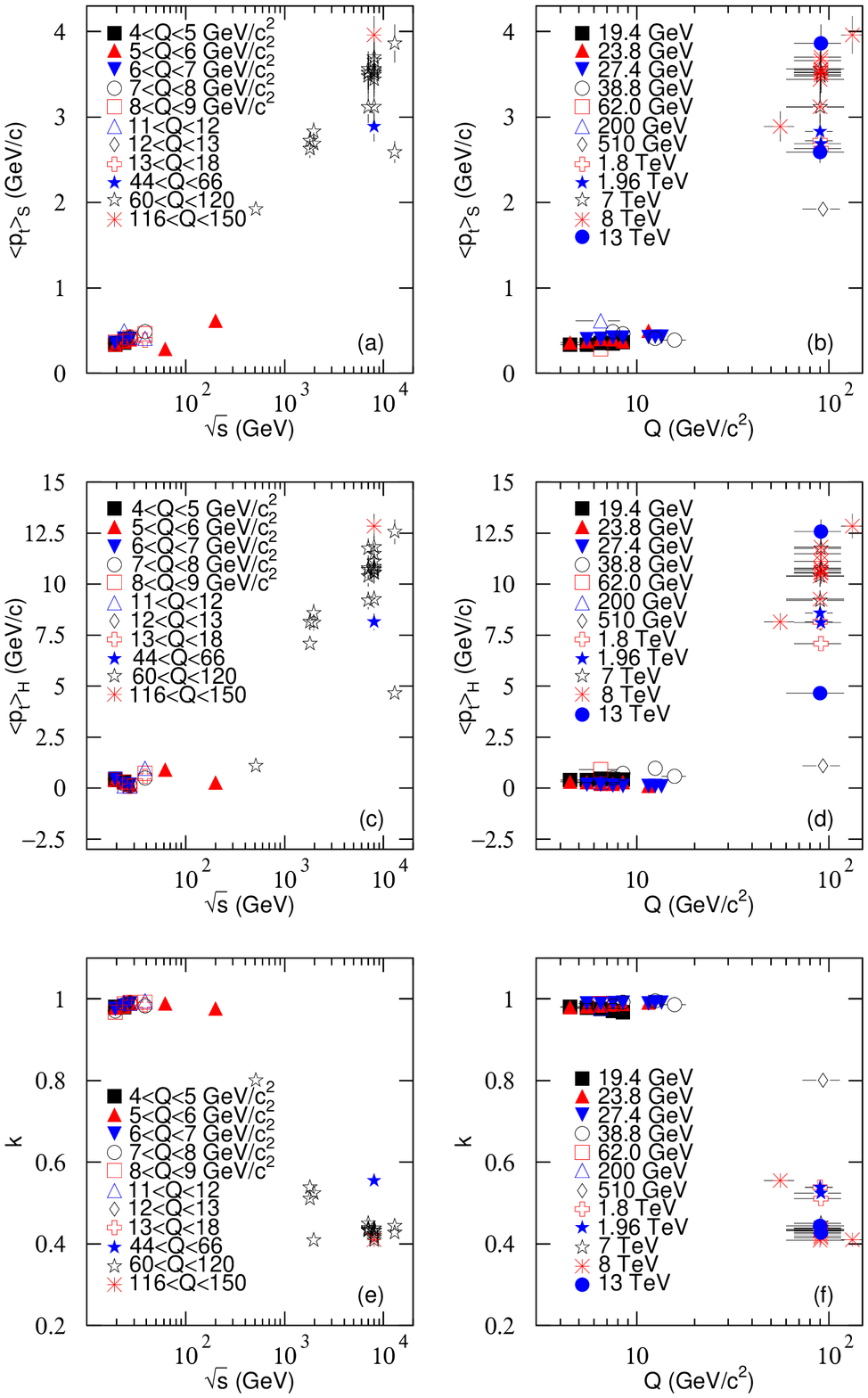}
\end{center}
{\small Figure 6: Similar to Figure 5, but showing the variation
of parameters $\langle p_{t}\rangle _{S}$ (a, b), $\langle
p_{t}\rangle _{H}$ (c, d), and $k$ (e, f) in the two-component
Erlang distribution with $\sqrt{s}$ (a, c, e) and $Q$ (b, d, f).}
\end{figure*}

\begin{figure*}[!htb]
\begin{center}
\includegraphics[width=12.0cm]{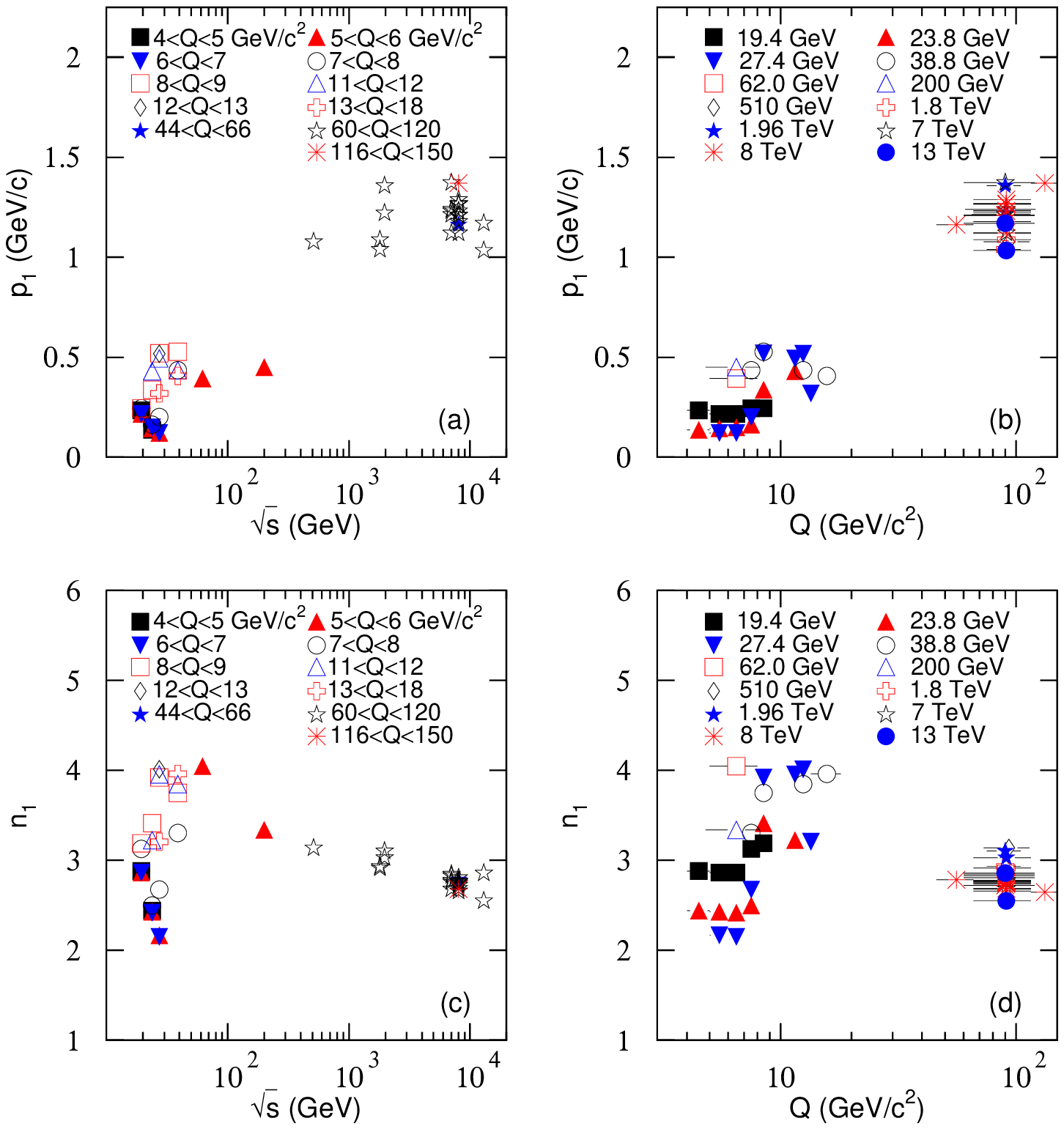}
\end{center}
{\small Figure 7: Similar to Figure 5, but showing the parameters
$p_{1}$ (a, b) and $n_{1}$ (c, d) in the Hagedorn function with
(a)(c) $\sqrt{s}$ and (b)(d) $Q$.}
\end{figure*}

{\section{Results and discussion}}

{\subsection{Comparison with data}}

Figure 1 shows the $p_T$ spectra of $\ell\bar \ell$ [with
different invariant masses ($Q$) or Feynman variables ($x_F$)]
produced by the Drell-Yan process in different collisions at
different energies (with different integral luminosities if
available in literature), where the concrete type of $\ell\bar
\ell$ is also given in each panel. The symbols $E$ and $\sigma$ on
the vertical axis denote the energy of $\ell\bar \ell$ and the
cross section of events respectively. Among them, the data points
presented in Figures 1(a)--1(c) are quoted from the proton-copper
($p$-Cu) collision experiments performed by the E288
Collaboration~\cite{E288}, and the collision energy per nucleon
pair ($\sqrt{s_{NN}}$ or $\sqrt{s}$ if in a simplified form) is
19.4, 23.8, and 27.4~GeV, respectively. The data points shown in
Figure 1(d) are the results of the $p$-Cu collision experiment
performed by the E605 Collaboration~\cite{38.8} at a collision
energy of 38.8~GeV. For the E288 Collaboration, the invariant mass
ranges from 4 to $14~\mathrm{GeV}/c^2$, while the corresponding
invariant mass ranges of the E605 Collaboration are from 7 to
$18~\mathrm{GeV}/c^2$. The experimental data points in
Figures~1(e) and 1(f) are from negative pions ($\pi^-$) induced
wolfram (W) ($\pi^-$-W) collisions at 21.7 GeV performed by the
FNAL-615 Collaboration~\cite{21.7}. The different symbols in
Figure 1(e) represent invariant mass $Q$ in the range of
4.05--$13.05~\mathrm{GeV}/c^2$ with different scalings, where the
units GeV/$c^2$ are not shown in the panel due to crowding space.
The Feynman variables range from 0 to 1 with a step of 0.1, as
shown in Figure 1(f). Different collaborations have different
intervals of $Q$ and $x_F$, while the detailed binning information
is marked in the panels. In some cases, the range of rapidity $y$
is not available due to other selection conditions such as the
complex polar coverages and sensitivities of detector components
or Feynman variable being used~\cite{21.7}. The total experimental
uncertainties are cited from refs.~\cite{E288,38.8,21.7} which
include both the statistical and systematic uncertainties if both
are available. The solid, dashed, and dotted curves in all panels
are the results of our fittings with the convolution of two
L\'{e}vy-Tsallis functions, the two-component Erlang distribution,
and the convolution of two Hagedorn functions, respectively. The
histograms in this and following figure correspond to QCD
calculations which will be discussed later. We use the
minimum-$\chi^2$ to evaluate the goodness of the fits, where
$\chi^2=\sum_j[({\rm Data}_j-{\rm Fit}_j)^2/{\rm
Uncertainty}_j^2]$ and $j$ is for the $j$-th data. We list the
results of the fits (parameters), the $\chi^2$, and the number of
degrees of freedom (ndof) in Table 1. The numbers $n_S=3$ and
$n_H=2$ which are not listed in the table to avoid trivialness.
One can see that the three functions can approximately describe
the $p_T$ spectra of $\ell\bar \ell$ produced by the Drell-Yan
process in high energy $p$-Cu and $\pi^-$-W collisions. The
two-component Erlang distribution describes better than the other
two functions.

Figure 2 shows the $p_T$ spectra of $\ell\bar \ell$ (with
different invariant mass $Q$) generated by the Drell-Yan process
in proton-proton ($p$-$p$ or $pp$) collisions and measured by
three different collaborations. The data points in Figure 2(a) are
from the experimental results measured by the R209
Collaboration~\cite{62a}. The collision energy is $\sqrt{s}=
62~\mathrm{GeV}$, and the $Q$ range is $5-8~\mathrm{GeV}/c^2$. The
data points in Figure 2(b) show the experimental results from the
PHENIX Collaboration~\cite{200}. The $\sqrt{s}$ is 200~GeV, the
$Q$ range is $4.8-8.2~\mathrm{GeV}/c^2$, and the rapidity range is
$1.2<y<2.2$. The data points in Figure 2(c) are from the
experimental results of the STAR Collaboration~\cite{510}. The
$\sqrt{s}$ is 510~GeV, the $Q$ range is $73-114~\mathrm{GeV}/c^2$,
and the rapidity range is $|y|<1$. In some cases, the range of
rapidity $y$ is not available due to other selection conditions
being used~\cite{62a}. The curves of the three fits are also shown
and the extracted parameters are presented in Table 2. One can see
that the three functions can approximately describe the $p_T$
spectra of $\ell\bar \ell$ produced by the Drell-Yan process in
high energy $pp$ collisions.

Similar to Figure 2, Figure~3 shows the $p_T$ spectra of $\ell\bar
\ell$ produced by the Drell-Yan process in proton-anti-proton
($p$-$\bar p$ or $p\bar p$) collisions with $66\leq Q<116$
GeV/$c^2$ (a, b), $75\leq Q<105$ GeV/$c^2$ (c), and $70\leq Q<110$
GeV/$c^2$ (d) at $\sqrt{s}= 1.8~\mathrm{TeV}$ (a, c) and
$\sqrt{s}= 1.96~\mathrm{TeV}$ (b, d). The data points in
Figures~3(a) and 3(b) are from the experiments of the
CDF~\cite{CDF1.8,CDF1.96} and D0
Collaborations~\cite{D01.8,D01.96-1,D01.96-2} respectively. In
some cases, the range of rapidity $y$ is not available due to
other selection conditions being used~\cite{CDF1.8,CDF1.96,D01.8}.
The curves of the three fits are also shown and the extracted
parameters are presented in Table 2. One can see that the three
functions can approximately describe the $p_T$ spectra of
$\ell\bar \ell$ produced by the Drell-Yan process in high energy
$p\bar p$ collisions.

Figure 4 shows the $p_T$ spectra of $\ell\bar \ell$ produced by
the Drell-Yan process in high energy $pp$ collisions, where the
data used in the figure are all examples and there is no bias
towards any specific CERN experiment. From Figures 4(a)--4(f), the
event samples [with different conditions ($\sqrt{s}$, $Q$, and
$y$)] are shown in the panels. The data points are quoted from the
experiments performed by the ATLAS
(a--d)~\cite{ATLAS7,ATLAS8,ATLAS13}, CMS (e)~\cite{CMS7,CMS8}, and
LHCb Collaborations (f)~\cite{LHCb7,LHCb8,LHCb13}. The extracted
fit parameters values are listed in Table~2. One can see that the
three functions can approximately describe the $p_T$ spectra of
$\ell\bar \ell$ produced by the Drell-Yan process in $pp$
collisions at ultrahigh energies.

From the above comparisons, we see that some of the data sets are
relatively poorly described by the fit. Notably, the FNAL-615
data, and the CDF data at 1.96 TeV, are not very well fitted.
Specifically, the fit of the CDF data returns a poor $\chi^2$,
unlike the other Tevatron experiment, as all LHC data sets have
reasonably good $\chi^2$. We would like to point out that the
relatively poor $\chi^2$ of some data is understandable due to the
fact that the fit function works well in most cases and does not
work well in a few cases. In addition, the data at high transverse
momentum has low statistics which causes large dispersion of the
data from the fit. In other cases, the low statistics happens to
be well fitted. To improve the relatively poor $\chi^2$ of some
fits, a better fit function could be used in the future, although
it is not feasible to find a fit function that works well in all
cases. One could improve the quality of the fits by concentrating
in the high-statistics regions or by rebinning the low-statistics
data.
\\

{\subsection{Tendency of parameters}}

In order to study the underlying law that governs the variation of
the extracted parameters with $\sqrt{s}$ and $Q$, we preset the
values of the fitted parameters as a function of the these two
quantities. Figures 5(a) and 5(b) show the trend of parameter $n$,
and Figures 5(c) and 5(d) show the trend of parameter $T$,
obtained by the fitting, using the L\'{e}vy-Tsallis function.
Comparing Figures 5(a) and 5(c), we can analyze the dependence of
parameters on $\sqrt{s}$. On average, it can be seen that as
$\sqrt{s}$ increases, the parameter $n$ increases quickly and then
decreases slowly, and the parameter $T$ increases slowly and then
significantly. There is a knee point for the trend of $n$ at
$\sqrt{s}\approx40$--50 GeV. Meanwhile, there is a boundary at
$\sqrt{s}\approx200$--500 GeV above which $T$ increases
significantly. Similarly, we compare Figures 5(b) and 5(d) and
analyze the variation of parameters with $Q$. It can be clearly
seen that the parameter $n$ increases firstly and then decreases,
and the parameter $T$ increases slowly and then significantly,
with the increase of $Q$. There is a knee point for the trend of
$n$ at $Q\approx14$--15 GeV/$c^2$. Meanwhile, there is a boundary
at $Q\approx20$--60 GeV/$c^2$ above which $T$ increases
significantly.

Figure~6 is similar to Figure~5, but shows the dependence of
parameters $\langle p_t\rangle_S$ (a, b), $\langle p_t\rangle_H$
(c, d), and $k$ (e, f) on $\sqrt{s}$ (a, c, e) and $Q$ (b, d, f)
obtained from the two-component Erlang distribution. One can see
that with increasing $\sqrt{s}$, $\langle p_t\rangle_S$ and
$\langle p_t\rangle_H$ increase slowly and then quickly, and $k$
decreases slowly and then quickly. For $\sqrt{s}$, there is a
boundary between 60--500 GeV for $\langle p_t\rangle_S$, between
500--1100 GeV for $\langle p_t\rangle_H$, and between 200--500 GeV
for $k$. Meanwhile, with increasing $Q$, $\langle p_t\rangle_S$
and $\langle p_t\rangle_H$ increase slowly and then quickly, and
$k$ decreases slowly and then quickly. There is a boundary at
$Q\approx20$--60 GeV/$c^2$.

In Figure 7, we show the parameters from the fits using the
Hagedorn function and their dependence on collision energy and
invariant mass. In Figures 7(a) and 7(b), one can see that the
parameter $p_1$ has a slight tendency to increase with the
increase of $\sqrt{s}$ and $Q$. In Figures 7(c) and 7(d), one can
see that the parameter $n_1$ decreases quickly and then slowly
with the increase of $\sqrt{s}$; $n_1$ also decreases slowly and
then quickly with increase of $Q$. There is a boundary for the
trend of $n_1$ at $\sqrt{s}\approx40$--200 GeV and
$Q\approx20$--50 GeV/$c^2$. We note that the variation of
parameter is obtained from the the average values of the extracted
parameters, for each $\sqrt{s}$ or $Q$.

It should be pointed out that the values of the parameters in
Figures 5--7 are all obtained by fitting the experimental data in
Figures 1--4 using the convolution of two L\'{e}vy-Tsallis
functions, the two-component Erlang distribution, and the
convolution of two Hagedorn functions, where the values obtained
from Figures 1(e) and 1(f) are not included. Firstly, this is
because the grouping of the quality in Figure 1(e) is different
from the grouping of other data, and there are already many other
groupings. Secondly, Figure 1(f) analyzes the $p_T$ spectra within
different ranges of Feynman variables, which is different from
others in terms of event sample. To avoid trivialness, we have not
put the fitting results of Figures 1(e) and 1(f) in Figures 5--7,
though these results are also shown in Table 1. We find that, by
analyzing these results, they do not contradict the trend of other
results presented in Figures 5--7.

The parameters $T$, $\langle p_t\rangle_S$, $\langle
p_t\rangle_H$, and $p_1$ show monotonous increasing trend when
$\sqrt{s}$ and $Q$ increase. Naturally, the variation degrees are
different. These increasing trends reflect that these parameters
describe the violent degree of collisions between two
(anti-)quarks in the Drell-Yan process. As the contribution
fraction of the first component in the two-component Erlang
distribution, $k$ decreasing with increasing $\sqrt{s}$ and $Q$
reflects naturally the increase of the contribution fraction of
the second component. The parameter $n$ increases firstly and then
decreases, and the parameter $n_1$ decreases, with the increase of
$\sqrt{s}$ and $Q$. This variation implies the change of
interaction pattern and strength. A possible explanation is that
the collision centrality between the two (anti-)quarks changes
from periphery to center, or the violent degree of collisions
increase, when $\sqrt{s}$ and $Q$ increase. We should pay more
attention on this variation in a future study.

In particular, the energy range considered in the present work is
enough for the formation of QGP. It is possible that the $p_T$
spectra of $\ell\bar \ell$ in Drell-Yan process are affected by
QGP. The non-monotonous change of $n$ implies the maximum
influence between Drell-Yan and QGP. According to the Tsallis
statistics, $n$ is opposite to the entropy index $q$ because
$n=1/(q-1)$. The maximum $n$ at $\sim40$ GeV implies that $q$ is
the closest to 1 at this energy. Meanwhile, at this energy, the
system is the closest to the equilibrium. The softest point
($\sim40$ GeV) of equation of state from the excitation function
of $q$ is consistent with our very recent work~\cite{59} in which
we stated that ``the onset energy of the partial phase transition
from hadron matter to QGP is 7.7 GeV and that of the whole phase
transition is 39 GeV" with the uncertainty of 1--2 GeV.

We explain further the softest point here. In the energy range
below $\sim40$ GeV, there is not enough volume for the
interactions between Drell-Yan and QGP due to partial phase
transition~\cite{59}. In the energy range above $\sim40$ GeV,
there is not enough time for the interactions between Drell-Yan
and QGP because the participants penetrate through each other
quickly. Insufficient volume and time result in the system being
not the closest to the equilibrium, though the system is still
close to equilibrium at the softest point, $Q\approx11$--18
GeV/$c^2$. At below (above) the softest point, we see naturally
smaller (larger) $Q$, except for $Q=5$--6 and 5--8 GeV/$c^2$ which
show below the softest point at 20--30 GeV and also show above the
softest point at 60--200 GeV, which should be studied in the
future.

At the energy below 200 GeV and the invariant mass below 20
GeV/$c^2$, the system is close to the onset energy of phase
transition; The parameters $T$, $\langle p_t\rangle_S$, $\langle
p_t\rangle_H$, $1-k$, and $p_1$ show similar small and almost flat
values respectively, and $n$ and $n_1$ show different trends. At
the energy of TeV and the invariant mass around 100 GeV/$c^2$, the
system is far away from the onset energy of phase transition; The
parameters $T$, $\langle p_t\rangle_S$, $\langle p_t\rangle_H$,
$1-k$, and $p_1$ show large values, and $n$ and $n_1$ show a trend
of decrease or saturation. These obvious characteristics in
parameters make a clear distinction between approaching to and far
away from the onset energy of phase transition from hadron matter
to QGP because QGP also affects the spectra of $\ell\bar \ell$ in
Drell-Yan process.
\\

{\subsection{Further discussion}}

In the exact factorization formula, the true TMD parton
distributions depend on the hard scale of the process (for the
Drell-Yan process, the invariant mass $Q$ of $\ell\bar \ell$) and
obey renormalization group equations. Although this scaling is not
included in the function $f_1(p_{t1})$ or $f_1(p_T-p_{t1})$ of Eq.
(6), it re-appears through the dependence of its parameters upon
$Q$ (and also the center-of-mass energy squared $s$). The
dependence of the parameters on $Q$ reflects the fact that the
function $f_1(p_{t1})$ or $f_1(p_T-p_{t1})$ must change its
parameters with the hard scale. This result is consistent with the
appropriate frameworks: collinear framework for large transverse
momentum, and TMD framework for small transverse
momentum~\cite{1a}.

\clearpage

\end{multicols}
\begin{sidewaystable}
%\begin{table*}
{\small Table 1: Values of parameters $n$ and $T$ in the
L{\'e}vy-Tsallis function (solid curves), $\langle
p_{t}\rangle_{S}$, $\langle p_{t}\rangle _{H}$, and $k$ in the
two-component Erlang distribution (dashed curves), as well as
$p_{1}$ and $n_{1}$ in the Hagedorn function (dotted curves) in
Figure 1. The $\chi^2$ is for each fit. The ndof is for the fits
of two-component Erlang distribution. For the fits of
L\'{e}vy-Tsallis function and Hagedorn function, the ndof needs
one more. For Figure 1(f), the selection is not the invariant mass
$Q$, but the Feynman variable $x_F$. $n_S=3$ and $n_H=2$ for
Figure 1, which are not listed in the table.} \vspace{-8mm}
{\scriptsize
\begin{center}
\begin{tabular}{cccccccccccccc}\\
\hline\hline
Figure & $\sqrt{s}$ (GeV) & $Q$ (GeV/$c^2$) & $n$ & $T$ (GeV) & $\chi^{2}$ & $\langle p_t\rangle_S$ (GeV/$c$) & $\langle p_t\rangle_H$ (GeV/$c$) & $k$ & $\chi^{2}$ & $p_1$ (GeV/$c$) & $n_1$ & $\chi^2$ & ndof \\
\hline
Figure 1(a)& $19.4$ & $4-5      $ & $3.910\pm 0.089$ & $0.161\pm 0.005$ & $ 122 $ & $0.294\pm 0.003$ & $0.611\pm 0.006$ & $0.961\pm 0.003$ & $ 52$ & $1.028\pm 0.006$ & $6.994\pm 0.022$ & $164$ & $ 10 $ \\
           & $    $ & $5-6      $ & $3.969\pm 0.089$ & $0.166\pm 0.005$ & $ 82  $ & $0.303\pm 0.004$ & $0.608\pm 0.006$ & $0.974\pm 0.003$ & $ 24$ & $1.031\pm 0.006$ & $6.992\pm 0.022$ & $ 99$ & $ 11 $ \\
           & $    $ & $6-7      $ & $4.008\pm 0.090$ & $0.169\pm 0.005$ & $ 34  $ & $0.327\pm 0.006$ & $0.588\pm 0.005$ & $0.978\pm 0.003$ & $ 10$ & $1.035\pm 0.006$ & $6.987\pm 0.022$ & $ 43$ & $ 10 $ \\
           & $    $ & $7-8      $ & $4.139\pm 0.093$ & $0.171\pm 0.005$ & $ 9   $ & $0.331\pm 0.008$ & $0.511\pm 0.004$ & $0.973\pm 0.003$ & $ 1 $ & $1.046\pm 0.006$ & $6.944\pm 0.021$ & $ 12$ & $ 8  $ \\
           & $    $ & $8-9      $ & $4.155\pm 0.098$ & $0.176\pm 0.005$ & $ 6   $ & $0.348\pm 0.008$ & $0.494\pm 0.004$ & $0.961\pm 0.003$ & $ 3 $ & $1.049\pm 0.006$ & $6.921\pm 0.020$ & $ 8 $ & $ 7  $ \\
\hline
Figure 1(b)& $23.8$ & $4-5      $ & $3.571\pm 0.087$ & $0.181\pm 0.006$ & $ 155 $ & $0.318\pm 0.004$ & $0.642\pm 0.006$ & $0.964\pm 0.003$ & $ 13$ & $1.002\pm 0.005$ & $6.264\pm 0.016$ & $157$ & $ 11 $ \\
           & $    $ & $5-6      $ & $3.892\pm 0.089$ & $0.182\pm 0.006$ & $ 96  $ & $0.339\pm 0.005$ & $0.598\pm 0.005$ & $0.967\pm 0.003$ & $ 22$ & $1.003\pm 0.004$ & $6.264\pm 0.015$ & $ 99$ & $ 11 $ \\
           & $    $ & $6-7      $ & $3.909\pm 0.092$ & $0.195\pm 0.007$ & $ 67  $ & $0.364\pm 0.007$ & $0.579\pm 0.005$ & $0.961\pm 0.003$ & $ 15$ & $1.005\pm 0.004$ & $6.230\pm 0.015$ & $ 74$ & $ 11 $ \\
           & $    $ & $7-8      $ & $3.988\pm 0.089$ & $0.196\pm 0.007$ & $ 38  $ & $0.379\pm 0.008$ & $0.543\pm 0.005$ & $0.970\pm 0.003$ & $ 14$ & $1.010\pm 0.004$ & $6.194\pm 0.013$ & $ 40$ & $ 10 $ \\
           & $    $ & $8-9      $ & $4.135\pm 0.102$ & $0.197\pm 0.007$ & $ 42  $ & $0.380\pm 0.008$ & $0.494\pm 0.004$ & $0.974\pm 0.003$ & $ 31$ & $1.011\pm 0.004$ & $6.193\pm 0.014$ & $ 42$ & $ 9  $ \\
           & $    $ & $11-12    $ & $4.410\pm 0.114$ & $0.241\pm 0.010$ & $ 3   $ & $0.427\pm 0.011$ & $0.433\pm 0.004$ & $0.982\pm 0.003$ & $ 3 $ & $1.025\pm 0.005$ & $6.021\pm 0.012$ & $ 4 $ & $ 6  $ \\
\hline
Figure 1(c)& $27.4$ & $5-6      $ & $3.264\pm 0.060$ & $0.187\pm 0.006$ & $ 103 $ & $0.367\pm 0.006$ & $0.773\pm 0.007$ & $0.947\pm 0.003$ & $ 35$ & $1.030\pm 0.005$ & $6.164\pm 0.010$ & $159$ & $ 11 $ \\
           & $    $ & $6-7      $ & $3.760\pm 0.084$ & $0.202\pm 0.008$ & $ 184 $ & $0.386\pm 0.008$ & $0.711\pm 0.006$ & $0.953\pm 0.003$ & $ 88$ & $1.041\pm 0.005$ & $6.144\pm 0.009$ & $180$ & $ 11 $ \\
           & $    $ & $7-8      $ & $3.653\pm 0.088$ & $0.213\pm 0.009$ & $ 144 $ & $0.403\pm 0.008$ & $0.688\pm 0.006$ & $0.953\pm 0.003$ & $ 38$ & $1.067\pm 0.005$ & $6.136\pm 0.010$ & $165$ & $ 11 $ \\
           & $    $ & $8-9      $ & $3.905\pm 0.092$ & $0.217\pm 0.009$ & $ 109 $ & $0.409\pm 0.008$ & $0.638\pm 0.006$ & $0.953\pm 0.003$ & $ 45$ & $1.082\pm 0.006$ & $6.125\pm 0.010$ & $122$ & $ 11 $ \\
           & $    $ & $11-12    $ & $4.247\pm 0.106$ & $0.230\pm 0.012$ & $ 21  $ & $0.442\pm 0.010$ & $0.602\pm 0.005$ & $0.931\pm 0.003$ & $ 10$ & $1.081\pm 0.005$ & $6.093\pm 0.009$ & $ 27$ & $ 11 $ \\
           & $    $ & $12-13    $ & $4.466\pm 0.113$ & $0.253\pm 0.014$ & $ 15  $ & $0.464\pm 0.011$ & $0.564\pm 0.005$ & $0.944\pm 0.003$ & $ 11$ & $1.086\pm 0.006$ & $6.089\pm 0.009$ & $ 16$ & $ 10 $ \\
           & $    $ & $13-14    $ & $4.986\pm 0.133$ & $0.256\pm 0.014$ & $ 4   $ & $0.468\pm 0.011$ & $0.552\pm 0.005$ & $0.944\pm 0.003$ & $ 3 $ & $1.084\pm 0.006$ & $6.092\pm 0.009$ & $ 4 $ & $ 8  $ \\
\hline
Figure 1(d)& $38.8$ & $7-8      $ & $3.211\pm 0.064$ & $0.254\pm 0.013$ & $ 19  $ & $0.446\pm 0.010$ & $0.654\pm 0.006$ & $0.983\pm 0.003$ & $ 6 $ & $1.257\pm 0.009$ & $6.191\pm 0.010$ & $ 42$ & $ 7  $ \\
           & $    $ & $8-9      $ & $4.371\pm 0.080$ & $0.262\pm 0.013$ & $ 47  $ & $0.469\pm 0.012$ & $0.602\pm 0.006$ & $0.992\pm 0.003$ & $ 11$ & $1.180\pm 0.008$ & $6.201\pm 0.011$ & $ 52$ & $ 10 $ \\
           & $    $ & $11.5-13.5$ & $4.694\pm 0.118$ & $0.267\pm 0.013$ & $ 32  $ & $0.473\pm 0.013$ & $0.529\pm 0.005$ & $0.995\pm 0.003$ & $ 7 $ & $1.194\pm 0.007$ & $6.190\pm 0.010$ & $ 44$ & $ 11 $ \\
           & $    $ & $13.5-18.0$ & $4.949\pm 0.131$ & $0.278\pm 0.015$ & $ 94  $ & $0.480\pm 0.013$ & $0.434\pm 0.007$ & $0.986\pm 0.003$ & $ 34$ & $1.197\pm 0.007$ & $6.171\pm 0.012$ & $ 98$ & $ 10 $ \\
\hline
Figure 1(e)& $21.7$ & $4.05-4.50  $ & $3.221\pm 0.085$ & $0.155\pm 0.005$ & $412$ & $0.397\pm 0.008$ & $0.473\pm 0.013$ & $0.958\pm 0.003$ & $107$ & $0.572\pm 0.006$ & $2.175\pm 0.012$ & $563$ & $13 $\\
           & $    $ & $4.50-4.95  $ & $3.073\pm 0.073$ & $0.163\pm 0.005$ & $227$ & $0.429\pm 0.009$ & $0.314\pm 0.005$ & $0.968\pm 0.003$ & $94 $ & $0.505\pm 0.005$ & $2.696\pm 0.010$ & $311$ & $14 $ \\
           & $    $ & $4.95-5.40  $ & $3.663\pm 0.089$ & $0.176\pm 0.006$ & $122$ & $0.411\pm 0.008$ & $0.481\pm 0.014$ & $0.966\pm 0.003$ & $88 $ & $0.512\pm 0.005$ & $2.357\pm 0.014$ & $180$ & $13 $ \\
           & $    $ & $5.40-5.85  $ & $3.348\pm 0.088$ & $0.165\pm 0.005$ & $47 $ & $0.425\pm 0.009$ & $0.632\pm 0.016$ & $0.955\pm 0.003$ & $57 $ & $0.595\pm 0.005$ & $2.803\pm 0.012$ & $79 $ & $11 $ \\
           & $    $ & $5.85-6.75  $ & $3.841\pm 0.112$ & $0.186\pm 0.008$ & $44 $ & $0.428\pm 0.009$ & $0.358\pm 0.006$ & $0.977\pm 0.003$ & $55 $ & $0.551\pm 0.005$ & $2.540\pm 0.014$ & $68 $ & $14 $ \\
           & $    $ & $6.75-7.65  $ & $3.453\pm 0.088$ & $0.188\pm 0.008$ & $32 $ & $0.442\pm 0.010$ & $0.499\pm 0.015$ & $0.971\pm 0.003$ & $48 $ & $0.563\pm 0.005$ & $2.746\pm 0.015$ & $70 $ & $11 $ \\
           & $    $ & $7.65-9.00  $ & $3.965\pm 0.090$ & $0.202\pm 0.008$ & $20 $ & $0.433\pm 0.010$ & $0.423\pm 0.009$ & $0.997\pm 0.003$ & $28 $ & $0.552\pm 0.005$ & $2.323\pm 0.014$ & $33 $ & $12 $ \\
           & $    $ & $9.00-10.35 $ & $3.321\pm 0.088$ & $0.167\pm 0.005$ & $15 $ & $0.379\pm 0.007$ & $0.482\pm 0.014$ & $0.983\pm 0.003$ & $10 $ & $0.506\pm 0.005$ & $2.904\pm 0.018$ & $20 $ & $11 $ \\
           & $    $ & $10.35-11.70$ & $3.013\pm 0.073$ & $0.172\pm 0.005$ & $17 $ & $0.372\pm 0.007$ & $0.219\pm 0.003$ & $0.984\pm 0.003$ & $13 $ & $0.561\pm 0.005$ & $2.767\pm 0.016$ & $20 $ & $9 $ \\
           & $    $ & $11.70-13.05$ & $2.707\pm 0.045$ & $0.153\pm 0.005$ & $3  $ & $0.354\pm 0.006$ & $1.491\pm 0.057$ & $0.979\pm 0.003$ & $3  $ & $0.521\pm 0.005$ & $3.312\pm 0.025$ & $3  $ & $7  $ \\
\hline
Figure 1(f)&  21.7  & $x_F$ ($1/c$) &                  &                  &       &                  &                  &                  &       &                  &                  &       & \\
           &        & $0.00-0.10  $ & $2.702\pm 0.045$ & $0.148\pm 0.005$ & $32 $ & $0.408\pm 0.008$ & $0.645\pm 0.016$ & $0.974\pm 0.003$ & $15 $ & $0.189\pm 0.005$ & $2.035\pm 0.008$ & $47 $ & $11 $ \\
           &        & $0.10-0.20  $ & $2.873\pm 0.048$ & $0.113\pm 0.004$ & $69 $ & $0.374\pm 0.007$ & $0.503\pm 0.015$ & $0.976\pm 0.003$ & $25 $ & $0.188\pm 0.005$ & $2.806\pm 0.012$ & $92 $ & $15 $ \\
           & $    $ & $0.20-0.30  $ & $3.978\pm 0.092$ & $0.187\pm 0.006$ & $90 $ & $0.386\pm 0.008$ & $0.487\pm 0.014$ & $0.979\pm 0.003$ & $46 $ & $0.190\pm 0.005$ & $2.865\pm 0.012$ & $134$ & $13 $ \\
           & $    $ & $0.30-0.40  $ & $2.938\pm 0.058$ & $0.136\pm 0.005$ & $120$ & $0.393\pm 0.008$ & $0.580\pm 0.015$ & $0.963\pm 0.003$ & $62 $ & $0.184\pm 0.005$ & $2.468\pm 0.011$ & $159$ & $13 $ \\
           & $    $ & $0.40-0.50  $ & $3.046\pm 0.061$ & $0.152\pm 0.005$ & $105$ & $0.393\pm 0.008$ & $0.512\pm 0.014$ & $0.975\pm 0.003$ & $57 $ & $0.185\pm 0.005$ & $3.581\pm 0.018$ & $190$ & $13 $ \\
           & $    $ & $0.50-0.60  $ & $2.820\pm 0.052$ & $0.158\pm 0.005$ & $106$ & $0.399\pm 0.008$ & $0.491\pm 0.014$ & $0.959\pm 0.003$ & $54 $ & $0.293\pm 0.010$ & $3.302\pm 0.016$ & $162$ & $13 $ \\
           & $    $ & $0.60-0.70  $ & $2.721\pm 0.045$ & $0.180\pm 0.006$ & $141$ & $0.383\pm 0.008$ & $0.403\pm 0.012$ & $0.969\pm 0.003$ & $52 $ & $0.298\pm 0.010$ & $2.841\pm 0.012$ & $198$ & $13 $ \\
           & $    $ & $0.70-0.80  $ & $2.938\pm 0.048$ & $0.158\pm 0.005$ & $119$ & $0.376\pm 0.007$ & $0.427\pm 0.013$ & $0.956\pm 0.003$ & $55 $ & $0.202\pm 0.008$ & $3.102\pm 0.016$ & $166$ & $13 $ \\
           & $    $ & $0.80-0.90  $ & $3.022\pm 0.063$ & $0.126\pm 0.004$ & $74 $ & $0.284\pm 0.004$ & $0.394\pm 0.011$ & $0.989\pm 0.003$ & $46 $ & $0.367\pm 0.012$ & $3.673\pm 0.018$ & $101$ & $11 $ \\
           & $    $ & $0.90-1.00  $ & $3.143\pm 0.068$ & $0.095\pm 0.012$ & $25 $ & $0.195\pm 0.002$ & $0.281\pm 0.008$ & $0.985\pm 0.003$ & $6  $ & $0.386\pm 0.015$ & $3.868\pm 0.022$ & $17 $ & $7  $ \\
\hline
\end{tabular}%
\end{center}}
%\end{table*}
\end{sidewaystable}

\newpage

\begin{sidewaystable}
%\begin{table*}
{\small Table 2: Values of parameters $n$ and $T$ in the
L{\'e}vy-Tsallis function (solid curves), $\langle
p_{t}\rangle_{S}$, $\langle p_{t}\rangle _{H}$, and $k$ in the
two-component Erlang distribution (dashed curves), as well as
$p_{1}$ and $n_{1}$ in the Hagedorn function (dotted curves) in
Figures 2--4. The $\chi^2$ is for each fit. The ndof is for the
fits of two-component Erlang distribution. For the fits of
L\'{e}vy-Tsallis function and Hagedorn function, the ndof needs
one more. For Figures 4(a) and 4(b), different $y$ ranges are
included. $n_S=3$ and $n_H=2$ for Figures 2 and 4, as well as
$n_S=2$ and $n_H=2$ for Figure 3, which are not listed in the
table.} \vspace{-8mm} {\scriptsize
\begin{center}
\begin{tabular}{cccccccccccccc}\\
\hline\hline
Figure & $\sqrt{s}$ (GeV) & $Q$ (GeV/$c^2$) & $n$ & $T$ (GeV) & $\chi^{2}$ & $\langle p_t\rangle_S$ (GeV/$c$) & $\langle p_t\rangle_H$ (GeV/$c$) & $k$ & $\chi^{2}$ & $p_1$ (GeV/$c$) & $n_1$ & $\chi^{2}$ & ndof \\
\hline
Figure 2(a)& $62.0$ & $5-8      $ & $3.887\pm 0.080$ & $0.136\pm 0.006$ & $21 $ & $0.287\pm 0.006$ & $0.913\pm 0.075$ & $0.989\pm 0.003$ & $26 $ & $1.001\pm 0.005$ & $6.829\pm 0.018$ & $18 $ & $9$\\
\hline
Figure 2(b)& $200 $ & $4.8-8.2  $ & $3.187\pm 0.068$ & $0.258\pm 0.015$ & $18 $ & $0.617\pm 0.017$ & $0.277\pm 0.010$ & $0.977\pm 0.003$ & $18 $ & $1.111\pm 0.005$ & $5.089\pm 0.016$ & $15 $ & $8$\\
\hline
Figure 2(c)& $510 $ & $73-114   $ & $3.162\pm 0.065$ & $0.792\pm 0.018$ & $4$   & $1.924\pm 0.062$ & $1.103\pm 0.110$ & $0.801\pm 0.003$ & $11 $ & $1.077\pm 0.012$ & $3.137\pm 0.019$ & $4  $ & $3$\\
\hline
Figure 3(a)& $1800$ & $66-116   $ & $2.618\pm 0.028$ & $0.820\pm 0.016$ & $25 $ & $2.631\pm 0.110$ & $7.092\pm 0.320$ & $0.511\pm 0.002$ & $24 $ & $1.088\pm 0.012$ & $2.916\pm 0.015$ & $37 $ & $33$\\
\hline
Figure 3(b)& $1960$ & $66-116   $ & $2.661\pm 0.018$ & $0.840\pm 0.018$ & $270$ & $2.690\pm 0.128$ & $8.120\pm 0.403$ & $0.524\pm 0.002$ & $134$ & $1.222\pm 0.013$ & $3.030\pm 0.016$ & $476$ & $53$\\
\hline
Figure 3(c)& $1800$ & $75-105   $ & $2.588\pm 0.021$ & $0.779\pm 0.016$ & $16 $ & $2.721\pm 0.101$ & $8.151\pm 0.400$ & $0.539\pm 0.002$ & $16 $ & $1.040\pm 0.012$ & $2.934\pm 0.015$ & $30 $ & $14$\\
\hline
Figure 3(d)& $1960$ & $70-110   $ & $2.745\pm 0.017$ & $0.904\pm 0.017$ & $14 $ & $2.830\pm 0.110$ & $8.580\pm 0.450$ & $0.539\pm 0.002$ & $13 $ & $1.359\pm 0.015$ & $3.106\pm 0.017$  & $26 $ & $9$\\
\hline
Figure 4(a)& $ 7000$& $66-116  $   & $              $ & $              $ & $   $ & $              $ & $              $ & $              $ & $   $ & $              $ & $              $ & $   $ & $ $\\
           & $   $  & $0\leq|y|<1$ & $2.415\pm 0.016$ & $0.996\pm 0.016$ & $3  $ & $3.521\pm 0.201$ & $10.710\pm0.501$ & $0.450\pm 0.002$ & $3  $ & $1.241\pm 0.013$ & $2.780\pm 0.015$ & $11 $ & $10$\\
           & $   $  & $1\leq|y|<2$ & $2.396\pm 0.016$ & $1.003\pm 0.016$ & $2  $ & $3.520\pm 0.201$ & $10.793\pm0.501$ & $0.439\pm 0.002$ & $3  $ & $1.231\pm 0.012$ & $2.760\pm 0.015$ & $7  $ & $10$\\
           & $   $  &$2\leq|y|<2.4$& $2.336\pm 0.012$ & $0.980\pm 0.016$ & $1  $ & $3.482\pm 0.200$ & $11.742\pm0.507$ & $0.434\pm 0.002$ & $1  $ & $1.121\pm 0.012$ & $2.679\pm 0.014$ & $1$   & $10$\\
\hline
Figure 4(b)& $8000$ & $66-116  $      & $              $ & $              $ & $   $ & $              $ & $              $ & $              $ & $   $ & $              $ & $              $ & $   $ & $   $\\
           & $ $    &$0.0\leq|y|<0.4$ & $2.385\pm 0.023$ & $1.023\pm 0.018$ & $2  $ & $3.520\pm 0.201$ & $10.571\pm0.505$ & $0.425\pm 0.002$ & $2$   & $1.232\pm 0.012$ & $2.738\pm 0.015$ & $6  $ & $7  $\\
           & $ $    &$0.4\leq|y|<0.8$ & $2.405\pm 0.023$ & $1.031\pm 0.020$ & $2  $ & $3.549\pm 0.203$ & $10.713\pm0.509$ & $0.432\pm 0.002$ & $2$   & $1.266\pm 0.013$ & $2.761\pm 0.016$ & $6  $ & $7  $\\
           & $ $    &$0.8\leq|y|<1.2$ & $2.411\pm 0.031$ & $1.031\pm 0.019$ & $2  $ & $3.561\pm 0.210$ & $11.760\pm0.515$ & $0.438\pm 0.002$ & $2$   & $1.267\pm 0.013$ & $2.767\pm 0.016$ & $6  $ & $7  $\\
           & $ $    &$1.2\leq|y|<1.6$ & $2.384\pm 0.023$ & $1.031\pm 0.020$ & $2  $ & $3.663\pm 0.213$ & $11.106\pm0.524$ & $0.438\pm 0.002$ & $1$   & $1.267\pm 0.013$ & $2.752\pm 0.016$ & $6  $ & $7  $\\
           & $ $    &$1.6\leq|y|<2.0$ & $2.298\pm 0.027$ & $0.990\pm 0.020$ & $1  $ & $3.700\pm 0.220$ & $11.817\pm0.528$ & $0.432\pm 0.002$ & $1$   & $1.178\pm 0.012$ & $2.686\pm 0.015$ & $3  $ & $7  $\\
           & $ $    &$2.0\leq|y|<2.4$ & $2.259\pm 0.022$ & $0.953\pm 0.020$ & $1  $ & $3.525\pm 0.200$ & $11.482\pm0.520$ & $0.419\pm 0.002$ & $1$   & $1.122\pm 0.010$ & $2.658\pm 0.015$ & $1  $ & $7  $\\
\hline
Figure 4(c)& $ $    & $46-66       $  & $2.918\pm 0.011$ & $0.987\pm 0.021$ & $24 $ & $2.890\pm 0.175$ & $8.152 \pm0.412$ & $0.555\pm 0.002$ & $18 $ & $1.164\pm 0.010$ & $2.787\pm 0.016$ & $50 $ & $7  $\\
           & $ $    & $66-116      $  & $2.402\pm 0.022$ & $1.038\pm 0.020$ & $4  $ & $3.493\pm 0.209$ & $10.540\pm0.505$ & $0.415\pm 0.002$ & $4  $ & $1.290\pm 0.012$ & $2.769\pm 0.015$ & $12 $ & $12 $\\
           & $ $    & $116-150     $  & $2.183\pm 0.011$ & $1.008\pm 0.011$ & $16 $ & $3.960\pm 0.221$ & $12.841\pm0.601$ & $0.410\pm 0.002$ & $10 $ & $1.371\pm 0.012$ & $2.684\pm 0.014$ & $40 $ & $7  $\\
\hline
Figure 4(d)&$13000$ & $66-116   $     & $2.177\pm 0.012$ & $0.962\pm 0.018$ & $11 $ & $3.862\pm 0.223$ & $12.570\pm0.603$ & $0.428\pm 0.002$ & $6  $ & $1.035\pm 0.010$ & $2.550\pm 0.014$ & $22 $ & $10 $\\
\hline
Figure 4(e)& $7000$ & $60-120 $ & $2.487\pm 0.020$ & $1.071\pm 0.020$ & $14 $ & $3.563\pm 0.205$ & $10.390\pm0.501$ & $0.433\pm 0.002$ & $5  $ & $1.373\pm 0.017$ & $2.831\pm 0.011$ & $24 $ & $6$\\
           & $8000$ & $       $ & $2.341\pm 0.024$ & $1.001\pm 0.019$ & $7  $ & $3.441\pm 0.203$ & $10.851\pm0.505$ & $0.409\pm 0.002$ & $19 $ & $1.210\pm 0.015$ & $2.719\pm 0.014$ & $11 $ & $6$\\
\hline
Figure 4(f)& $7000 $& $60-120 $ & $2.483\pm 0.011$ & $0.936\pm 0.017$ & $19 $ & $3.114\pm 0.183$ & $9.211\pm 0.432$ & $0.436\pm 0.002$ & $32 $ & $1.214\pm 0.011$ & $2.846\pm 0.015$ & $36 $ & $8$\\
           & $8000 $& $       $ & $3.438\pm 0.010$ & $0.941\pm 0.017$ & $15 $ & $3.121\pm 0.183$ & $9.263\pm 0.431$ & $0.433\pm 0.002$ & $30 $ & $1.211\pm 0.011$ & $2.810\pm 0.015$ & $47 $ & $8$\\
           & $13000$& $       $ & $3.394\pm 0.045$ & $0.996\pm 0.020$ & $17 $ & $2.590\pm 0.129$ & $4.653\pm 0.209$ & $0.444\pm 0.002$ & $36 $ & $1.170\pm 0.010$ & $2.859\pm 0.015$ & $16 $ & $8$\\
\hline
\end{tabular}%
\end{center}}
%\end{table*}
\end{sidewaystable}

\begin{multicols}{2}

In addition, the formulae for the evolution of TMD parton
distributions are rather complicated and the resulting effect is
difficult to parameterize with simple expressions. The present
work is a preliminary attempt for the purpose of parameterization
with simple expressions. Meanwhile, as a preliminary attempt, the
present work is not accurate in some cases which show large
$\chi^2$/ndof. In other cases, $\chi^2$/ndof is approximately 1 or
not too large. In any case, we firmly believe that the underlying
physics law is knowable and simple. A very complicated expression
for the transverse momentum spectra of $\ell\bar \ell$ in the
Drell-Yan process in high energy collisions is not the final one.
More work for simplifying the expressions are needed in the
future.

After describing the spectra of $\ell\bar \ell$ in Drell-Yan
process, it is possible to subtract the contribution of Drell-Yan
process from the spectra of $\ell\bar \ell$ in final state and
leave behind only the contribution of QGP. Then, one may study the
excitation function of related parameters from the spectra of
$\ell\bar \ell$ contributed purely in QGP conditions and search
for the softest points of equation of state from the excitation
function. The energies corresponding to the softest points are
expected to connect with the critical energy of phase transition
from hadron matter to QGP. If both the spectra of $\ell\bar \ell$
in Drell-Yan process and in QGP degeneration process are described
by simple functions, one can study the phase transition more
conveniently. We hope that the present work is significant in
methodology for the study in the future.

We now discuss the results as they pertain to QCD
calculations~\cite{20,21,62a,ATLAS13,70,71,72}. To study more
deeply, and in a visual way, the connection between our formalism
and the standard QCD resummation, we present a direct comparison
in Figures 1--4 as examples. As can be seen, the histograms
presented in Figures 1--4 represent the results with different
treatments based on QCD calculations, which will be discussed
separately in the following.

The histograms in Figures 1(a)--1(d) are directly quoted from
those in ref.~\cite{20} in which the next-to-next-to-leading order
(NNLO) calculation is performed for nonperturbative structure of
semi-inclusive deep-inelastic scattering and the Drell-Yan
scattering at small transverse momentum within TMD factorization.
The histograms in Figures 1(e) and 1(f) are directly quoted from
those in ref.~\cite{70} in which the NNLO calculation is performed
for the Drell-Yan processes within TMD factorization. In the
figure, some histograms are re-normalized to the data to give a
better comparison.

The histogram in Figure 2(a) is transformed from the curve in
ref.~\cite{62a} to fit the style of other panels and to give a
clear display. In the transformation from the curve to histogram,
the areas under the histogram and curve in a given transverse
momentum bin are kept being the same. In ref.~\cite{62a}, the
calculation of QCD convoluted with Gaussian function is used. The
histogram in Figure 2(b) is directly quoted from that in
ref.~\cite{71} in which the transverse momentum spectrum of low
invariant mass Drell-Yan is produced at next-to-leading order
(NLO) in the parton branching method. The histogram in Figure 2(c)
is directly quoted from that in ref.~\cite{21} in which the TMD
parton distributions are considered up to
next-to-next-to-next-to-leading order logarithmic (N$^3$LO or
N$^3$LL) from the Drell-Yan data.

The meanings of histograms in Figures 3, 4(a), 4(b), 4(c) with
$46\leq Q<66$ and $116\leq Q<150$ GeV/$c^2$, 4(e), and 4(f) are
the same as Figures 1(a)--1(d), which will not be discussed
anymore. The histogram in Figure 4(c) with $66\leq Q<116$
GeV/$c^2$ is directly quoted from that in ref.~\cite{72} in which
the calculation is performed due to the Drell-Yan transverse
momentum resummation of fiducial power corrections at N$^3$LL. The
histograms in Figure 4(d) are directly quoted from those in
ref.~\cite{ATLAS13} in which the transverse momentum distributions
of the Drell-Yan $\ell\bar \ell$ are calculated up to
NNLO+N$^3$LL.

From the above descriptions, one can see that the formalism of
this paper is flexible in the fit to data. We may compare the fits
with more QCD predictions, it does not matter with or without the
TMD PDFs. Except for several papers in which the Drell-Yan
transverse momentum spectrum is predicted based on QCD
factorization are referenced in the introduction~\cite{20,21} and
the above discussion~\cite{62a,ATLAS13,70,71,72}, more works
related to the PDFs or TMDs in QCD are available
recently~\cite{60,61,62,63,64,65,66,67,68,69}. These QCD
factorizations are complex in the calculation and show different
forms of formalization from this paper but result in similar
shapes for the dilepton spectra as observed in experiments. By
adjusting the parameters, the formalism of this paper can flexibly
fit the QCD predictions with or without the TMD PDFs. Compared to
predictions from collinear PDFs, the formalism of this paper is
closer to TMD PDFs due to the flexible parameter selection in the
fit. The PDFs or TMDs in QCD and other QCD-based analyses reveal
the dynamic aspect of particle production process. The formalism
used by us reflects the statistical behavior of the produced
particles.
\\

{\section{Summary and conclusions}}

We have studied the transverse momentum spectra of lepton pairs
generated by the Drell-Yan process in $p$-Cu, $\pi^-$-W, and $pp$
($p\bar p$) collisions over an energy range from $\sim20$ GeV to
above 10 TeV. The low energy data come from the E288, E605, R209,
PHENIX, and STAR Collaborations. The high energy data come from
the CDF, D0, ATLAS, CMS, and LHCb Collaborations. The invariant
mass range of the final-state particles produced in the collisions
also has a large span of $4<Q<150~\mathrm{GeV}/c^2$. Three types
of probability density functions are used to fit and analyze the
collected experimental data. All the three functions are
approximately in agreement with the experimental data and the QCD
calculations. Some parameters are obtained.

In the convolution of two L\'{e}vy-Tsallis functions, as
increasing $\sqrt{s}$, there is a knee point for the trend of $n$
at $\sqrt{s}\approx40$--50 GeV. Meanwhile, there is a boundary at
$\sqrt{s}\approx200$--500 GeV above which $T$ increases
significantly. With the increase of $Q$, there is a knee point for
the trend of $n$ at $Q\approx14$--15 GeV/$c^2$. Meanwhile, there
is a boundary at $Q\approx20$--60 GeV/$c^2$ above which $T$
increases significantly. In the two-component Erlang distribution,
there is a boundary at $\sqrt{s}\approx60$--500 GeV for $\langle
p_t\rangle_S$, 500--1100 GeV for $\langle p_t\rangle_H$, and
200--500 GeV for $k$, above which $\langle p_t\rangle_S$, $\langle
p_t\rangle_H$, and $1-k$ increase quickly. Meanwhile, there is a
boundary at $Q\approx20$--60 GeV/$c^2$ above which $\langle
p_t\rangle_S$, $\langle p_t\rangle_H$, and $1-k$ increase quickly.
In the convolution of two Hagedorn functions, $p_1$ increases
obviously with increasing $\sqrt{s}$ and $Q$. There is a boundary
for the trend of $n_1$ at $\sqrt{s}\approx40$--200 GeV and
$Q\approx20$--50 GeV/$c^2$.

With increasing $\sqrt{s}$ and $Q$, the parameters $T$, $\langle
p_t\rangle_S$, $\langle p_t\rangle_H$, $1-k$, and $p_1$ show
monotonous increasing trend. These increasing trends reflect that
these parameters describe the violent degree of collisions between
quark and anti-quark in the Drell-Yan process. The parameter $n$
increases initially and then decreases, and the parameter $n_1$
decreases, with the increase of $\sqrt{s}$ and $Q$. This variation
implies the change of interaction pattern and strength. A possible
explanation is that the collision centrality between quark and
anti-quark changes from periphery to center, or the violent degree
of collisions increase, when $\sqrt{s}$ and $Q$ increase. The fit
results in this paper are comparable to the QCD NNLO and N$^3$LL
results with TMD PDFs.

In conclusion, the large number of events collected by the
investigated experiments allows us to study the statistical
behavior of dilepton production in hadron and nuclei collisions.
The lepton pairs can be produced through the Drell-Yan process and
the QGP degeneration process. Both processes are predicted by QCD
and are described well enough by our statistical thermodynamics
models, especially in kinematic regions with sufficiently large
numbers of events.
\\
\\
{\bf Data Availability}

The data used to support the findings of this study are included
within the article and are cited at relevant places within the
text as references.
\\
\\
{\bf Ethical Approval}

The authors declare that they are in compliance with ethical
standards regarding the content of this paper.
\\
\\
{\bf Disclosure}

The funding agencies have no role in the design of the study; in
the collection, analysis, or interpretation of the data; in the
writing of the manuscript, or in the decision to publish the
results.
\\
\\
{\bf Conflicts of Interest}

The authors declare that there are no conflicts of interest
regarding the publication of this paper.
\\
\\
{\bf Acknowledgments}

X.-H. Zhang and F.-H. Liu's work was supported by the National
Natural Science Foundation of China under Grant Nos. 12047571,
11575103, and 11947418, the Scientific and Technological
Innovation Programs of Higher Education Institutions in Shanxi
(STIP) under Grant No. 201802017, the Shanxi Provincial Natural
Science Foundation under Grant No. 201901D111043, and the Fund for
Shanxi ``1331 Project" Key Subjects Construction.
\\

\end{multicols}
\end{document}